\documentstyle[12pt]{article}

\newcommand{\be}{\begin{equation}}
\newcommand{\ee}{\end{equation}}
\newcommand{\bea}{\begin{eqnarray}}
\newcommand{\eea}{\end{eqnarray}}
\newcommand{\bd}{\begin{displaymath}}
\newcommand{\ed}{\end{displaymath}}
\newcommand{\bi}{\begin{itemize}}
\newcommand{\ei}{\end{itemize}}
\newcommand{\bc}{\begin{center}}
\newcommand{\ec}{\end{center}}
\newcommand{\bfl}{\begin{flushleft}}
\newcommand{\efl}{\end{flushleft}}
\newcommand{\bfr}{\begin{flushright}}
\newcommand{\efr}{\end{flushright}}
\newcommand{\f}{\frac}

\def\6{\partial}

\def\={\!\!\!&=&\!\!\!}
\def\+{\!\!\!&&\!\!\!+~}
\def\-{\!\!\!&&\!\!\!-~}

\renewcommand{\AA}{{\cal A}}

\newcommand{\FF}{{\cal F}}

\begin{document}

\title{Duality between coordinates and wave functions on noncommutative 
space}	
\author{Ion V. Vancea\footnote{ion@dfm.ffclrp.usp.br}}
\maketitle

\begin{center}
{\small 
Departamento de F\'{\i}sica e Matem\'{a}tica,\\ Faculdade de Filosofia, 
Ci\^{e}cias e Letras de Ribeir\~{a}o Preto 
USP,\\ Av. Bandeirantes 3900, Ribeir\~{a}o Preto 14040-901, SP, Brasil}
\end{center}

\begin{abstract}
The relation between coordinates and the solutions of the stationary 
Schr\"{o}dinger equation
in the noncommutative algebra of functions on $R^{2N}$ is discussed. We 
derive this relation for a certain class of wave functions for which 
the quantum prepotentials depend linearly on the coordinates similarly to 
the commutative case. Also, the differential equation satisfied by the 
prepotentials is given.

\end{abstract}

\newpage

Recently, it has been noted in \cite{fm1,fm2} that by introducing a 
functional on the space of the solutions of the stationary Schr\"{o}dinger 
equation (and also the Klein-Gordon equation) which connects two 
linearly independent solutions and is called a quantum prepotential, one can 
express a space-time coordinate as function of the corresponding 
solutions of the Schr\"{o}dinger equation and the prepotential. This method 
works well for any differential equation of second order with constant 
coefficients and constant Wronskian and also for a wide class of partial 
differential equations of second order with constant coefficients. 
Tentatives to locally generalize this construction  to gravity and to the 
Dirac equation have been made in \cite{ivv1,ivv2,ivv3} but it has not 
been found the same simple and clear structure as in the case of the 
second order equations. An inspection of the relations involved in the 
duality between the coordinates and the wave-functions shows that the 
duality is actually a Legendre transformation and suggests the physics 
should not change if one passes from one side of the duality to the other 
side. In a more concrete fashion, it has been postulated that systems 
with an arbitrary potential, if subjected to an Equivalence Postulate 
applied to the Hamilton-Jacobi equation, manifest an invariance under the 
Legendre transformation between the coordinates and momenta 
\cite{fm3,fm4,fm5,fm6,fm7,fm8,fm9,fm10,fm11}. (For earlier related works see 
\cite{fl1,fl2,fl3,fl4,ca1,ca2,ca3}.) This shows that the duality between the 
coordinates and wave functions possibly could be related to more deeper 
properties of the physical systems. On the other hand, it allows us to 
express the coordinates of space-time in terms of quantum quantities, 
which represents a new approach to the structure of space-time at high 
energies. From this point of view, the parametrization of space-time in 
terms of real numbers is secondary since it comes out of relations 
between prepotentials and wave functions which are quantum 
and thus more fundamental. This interpretation raises a series of 
interesting questions concerning the symmetries of space-time, the structure 
of space-time at the planckian scale, the properties of the quantum 
prepotentials and their form in other representations, and so on. We are 
going to address some of this questions in future works.

The aim of the present letter is to derive the duality between 
coordinates and wave functions at the Planck scale in the framework of the 
noncommutative space-time. Our motivation relies on the observation that 
the structure of the space-time at higher energies should be different 
from the classical picture only at the Planck scale where the gravity has 
the same strength as the other fundamental interactions. However, the 
best understanding of the fundamental interactions at this scale is 
presently given by the string theory. In a more general setting, the string 
theory induces a noncommutative structure of the space-time. This is 
due to the presence of the Kalb-Ramond field from the closed string 
spectra. The noncommutative structure is tractable in the particular case 
where the Kalb-Ramond field is constant \cite{sw}. Since in the limit of 
infinite string tension the strings behave as particles we expect to 
recover the particle Quantum Mechanics at some stage in this limit. This 
justifies the study of Quantum Mechanics on noncommutative spaces and 
sets out the framework of this study.

The noncommutative structure that emerges from the string theory is 
given by the following product \cite{sw} between the smooth functions on 
the Euclidean $R^{2N}$
\begin{equation}
\left( f\ast g \right) \left( x \right) =e^{\frac{i}{2}\theta 
^{mn}\partial
_{m}^{\left( 1\right) }\partial _{n}^{\left( 2\right) }}f\left( 
x_{1}\right)
g\left( x_{2}\right) \left| _{x_{1}=x_{2}=x}^{{}}\right. .
\label{starproduct}
\end{equation}
Here, $\theta^{mn}=-\theta^{nm}$ are of order of $L^2$, where $L$ is 
the fundamental length of the problem. In what follows we take $L=l_P$, 
the Planck length, according to the above discussion. The relation 
(\ref{starproduct}) can be interpreted as a map between the algebra of 
commutative smooth functions $\FF (R^N )$ on $R^N$ denoted by 
$\AA_{\bullet}$ and some infinite dimensional noncommutative associative algebra 
$\AA_{\star}$. The map (\ref{starproduct}) associates to each element $f 
\in \AA_{\bullet }$ an element denoted by the same symbol in 
$\AA_{\star}$ and called a {\em noncommutative function} on $R^{2N}$ and allows us 
to interpret $\AA_{\star}$ as a deformation of $\AA_{\bullet}$ with the 
noncommutative parameters $\theta^{mn}$. The two interpretations of the 
relation (\ref{starproduct}), i. e. as a new product rule on the space 
of smooth functions and as a map between two algebras, are both useful 
to study different aspects of noncommutativity of the space-time. 

Now let us define our general setting. The string theory describes the 
fundamental interactions at the Planck scale. Therefore, in this paper 
we consider the string length fixed to the Planck length. However, in 
infinite tension limit, the string length goes to zero. Thus, the Planck 
length also goes to zero, which is the limit of Quantum Mechanics. 
Consequently, the essential noncommutative contribution to the star product 
(\ref{starproduct}) comes from the linear terms in $\theta^{mn}$'s. 
This fixes our approximation. In particular, one can easily check out that 
(\ref{starproduct}) implies
\be
\left[ x^m , x^n \right]_{\star} = i \theta^{mn}.
\label{ncx}
\ee
The above relation is valid not only in the linear approximation of the 
star product, but also in its general form. If we treat $x^m$'s as 
cordinates on the noncommutative $R^{2N}$ manifold, their noncommutativity 
raises the question of the corresponding induced structure in the 
tangent and the cotangent spaces (see, for example \cite{cz,siv} for recent 
results on geometry with noncommutative parameters). More important for 
us, it posses the problem of choosing either noncommutative or 
commutative linear momenta associated with $x^m$'s. As discussed in
\cite{cs,ca,rpm,rsv,pah,baner}, the noncommutative momenta lead to 
usual commutative systems in the presence of a magnetic field while a 
genuine noncommutative situation is obtained if one considers commuting 
momenta. This remark, together with the quantization procedure, justifies 
working with the following algebra
\bea
\left[ x^m, p_n \right]_{\star} &=& i \hbar \delta^{m}_{n},
\label{quantrel}\\
\left[ p_m , p_n \right]_{\star} &=& 0.
\label{momrel}
\eea
Together with the relation (\ref{starproduct}), the commutator 
(\ref{momrel}) states that the linear momenta are independent of $x^m$'s. The 
stationary Schr\"{o}dinger equation is constructed by analogy with the 
commutative case and has the following form \cite{lm}
\be
H(x,p)\star\Psi (x) = E \Psi(x).
\label{scheq}
\ee
Here, the wave-function $\Psi$ belongs to the noncommutative algebra 
$\AA_{\star}$. It is important to note that the equation (\ref{scheq}) 
should be interpreted as a formal algebraic equation since the 
noncommutative coordinates do not form a complete set of observables. The analogy 
with the commutative case will be our guide in the search of a duality 
between the noncommutative coordinates (\ref{ncx}) and the solutions of 
the equation (\ref{scheq}) in the algebra $\AA_{\star}$. This duality 
represents an extension of the similar result from the commutative case. 
Also, it is an interesting mathematical problem by itself. 

Consider the explicit form of the Schr\"{o}dinger equation 
(\ref{scheq}) 
\be
\left[ - \f{\hbar^2}{2M}\sum_{m=1}^{2N}\6^{2}_{m} + V \right] \star 
\Psi = E \Psi ,
\label{scheqexpl}
\ee 
where $V(x)$ is an arbitrary function from $\AA_{\star}$ and $M$ is the 
mass of particle. As can be easily checked out, the star product in the 
kinetic term is equal to the commutative product. Following the 
commutative case \cite{fm1}, we treat the coordinate $x^k$ for $k=1, 2, \ldots 
, 2N$ as a variable and fix the coordinates $x^s$ for $s \neq k$. Thus, 
the equation (\ref{scheqexpl}) splits in to a system of $2N$ dependent 
differential equations of second order in the algebra $\AA_{\star}$. 
The coordinates $x^s$ play the role of paramenters in each of these 
equations. Since the inverse of any nonvanishing function is well defined in 
the noncommutative algebra $\AA_{\star}$ being equal to its inverse in 
the commutative algebra, one can write the system from 
(\ref{scheqexpl}) under the following form
\be
\left( -\f{\hbar^2}{2M}\6^{2}_{k} + V_{k} \star \Psi \right)( x ) = 
E\Psi (x),
\label{eqsyst}
\ee
and the potential in the $x^k$ direction has form
\be
V_{k} (x) = V(x) - \f{\hbar^2}{2M} \sum_{s \neq k}^{2N} \left( 
\6^{2}_{s} \Psi \star \Psi^{-1} \right) (x),
\label{systempot}
\ee
for all $k$. In order to establish a duality between the coordinates 
and the wave functions, we need two linearly independent solutions of the 
equation (\ref{eqsyst}). In the commutative case, the existence of them 
is guaranteed by theorems concerning the Wronskian function which are 
not always available in the noncommutative algebra. In fact, as on can 
see from (\ref{starproduct}), the Schr\"{o}dinger equation contains 
derivatives of arbitrary large order. Even if we limit ourselves to the 
first order in $\theta$'s, the corresponding equation is of second order 
with nonvanishing coefficient of the first order derivatives. Therefore, 
our problem is to find the conditions for which the solutions of the 
equation (\ref{eqsyst}) admit a dual representation in terms of 
coordinates. 

Let us assume that there are two solutions of the equation 
(\ref{eqsyst}) denoted by $\Psi_k$ and $\tilde{\Psi}_k$. If they are linearly 
dependent, i. e. there are two nonzero complex numbers $c_k$ and 
$\tilde{c_k}$ such that the relations
\be
\Psi_k = - \frac{\tilde c_k}{c_k}\tilde{\Psi}_k~~,~~\6_k \Psi_k = - 
\frac{\tilde c_k}{c_k}\6_k\tilde{\Psi}_k
\label{lindep}
\ee
hold simultaneously, then the following function
\be
W^{k}_{NC} = \6_k \Psi_k \star \tilde{\Psi}_k - \6_k \tilde{\Psi}_k 
\star \Psi_k ,
\label{ncwronsk}
\ee
is zero at each point $x \in R^{2N}$. This suggest using the 
noncommutative wronskian as defined by the above relation as a criterion for the 
behaviour of the two solutions. Assume now that the two solutions of 
the equation (\ref{eqsyst}) are linearly independent. The derivative of 
$W^{k}_{NC}$ with respect to the variable $x^k$ follows from the 
Leibniz' property
of the derivative in $\AA_{\star}$ and has the following form
\be
\6_k W^{k}_{NC} (x) = \left[ \6_k \Psi_k , \6_k \tilde{\Psi}_k 
\right]_{\star} (x)
+ \{ \left( V_k - E \right)\star \left[ \tilde{\Psi}_k , \Psi_k 
\right]_{\star} \} (x).
\label{derivwronsk}
\ee
Here and in what follows we are taking $\hbar^2 / 2M = 1$.
Next, let us introduce the quantum prepotential defined as in the 
commutative case \cite{fm1} by the following relation
\be
\tilde{\Psi}_k \equiv \frac{\6 \FF^{k} \left[ \Psi_k \right]}{\6 
\Psi_k}.
\label{defF}
\ee
The quantum prepotential $\FF^k$ is a functional on the space of the 
solutions of the equation (\ref{eqsyst}) for each $k$. In order to 
compute its derivative with respect to the coordinate $x^k$, one has to solve 
the ambiguity of ordering the terms in the chain product. Indeed, one 
has two nonequivalent possibilities for that, namely the left- and 
right-derivatives given by the following relations
\be
\6_k \FF^{k}_L \equiv \frac{\6 \FF^k}{\6_k \Psi_k} \star \6_k \Psi_k 
~~,~~
\6_k \FF^{k}_R \equiv \6_k\Psi_k \star \frac{\6 \FF^k}{\6_k \Psi_k}.
\label{leftrightderiv}
\ee
It turns out that the symmetriezed derivative defined by 
\be
\6_{k}^{S} \FF^k = \f{1}{2}\left( \6_k \FF^{k}_L + \6_k\FF^{k}_R 
\right)
\label{symmder}
\ee
satisfies the following relation
\be
2 \6_k^{S} \FF^k = \6_k \left( \tilde{\Psi}_k \star \Psi_k \right) + 
W^{k}_{NC} ,
\label{Frel}
\ee
which is similar to the one obtained in the commutative space 
\cite{fm1}. Also, if one uses the symmetrized derivative (\ref{symmder}) one can 
show that in the first order in $\theta$'s 
\be
\6_k^{S} \FF^k  = \left. \6_k \FF^{k} \right|_{\AA_{\bullet}}.
\label{derfirst}
\ee
The above relation allows us to define the differential of the quantum 
prepotential as follows
\be
d \FF^{k} \equiv \sum_{k=1}^{2N} \6_k^{S} \FF^{k} dx^k .
\label{diffF}
\ee
It is straightforward to integrate out the relation (\ref{Frel}). 
However, the result depends on the explicit form of the noncommutative 
wronskian which, in general, is not a constant function of $x^k$. 
Nevertheless, a similar result to the commutative case can be achieved by 
imposing that $W^{k}_{NC}$ satisfies 
\be
\theta^{mn} \{ \left( V_k -E \right) \6_m \Psi_k \6_n \tilde{\Psi}_k + 
\6_m \6_k \Psi_k \6_n \6_k \tilde{\Psi}_k \} = 0 .
\label{constW}
\ee
The above relation represents a constraint to obtain a linear equation 
in $x^k$ after integration. If one has in the r.h.s. of the equation a 
different function on $x^k$, one has a different dependence of the 
coordinate $x^k$ on the corresponding wave function and  prepotential.  
Just for completeness, let us give the expression of the noncommutative 
wronskian to the first order in $\theta$'s
\be
W^{k}_{NC} (x) = W^{k}(x) + \f{i}{2} \theta^{mn} \left[ \f{\6^2 
\FF^k}{\left(\6 \Psi_k \right)^2} (x) - \f{\6^3 \FF^k }{\left( \6 \Psi_k 
\right)^3 } (x) \right] \6_m \6_k \Psi_k (x) \6_n \Psi_k (x).
\label{Wncfirstord}
\ee
Here, $W^k$ represents the commutative wronskian.
If the equation (\ref{constW}) is satisfied, the relation between the 
noncommutative coordinate $x^k$ and the wave function has the following 
form
\be
x^k = \FF^k \left[ \Psi_k \right] - \f{1}{2} \tilde{\Psi}_k \star 
\Psi_k - f^k (x^s),
\label{xpsi}
\ee
where $f^k (x^s)$ is a function of the parameters $x^s$, $ s\neq k$ 
only. The noncommutative algebra (\ref{ncx}) and (\ref{quantrel}) imposes 
further constraints on the relation between the noncommutative 
potential and the wave function
\bea
\left[ \FF^k \left[ \Psi_k \right] - \f{1}{2} \tilde{\Psi}_k \star 
\Psi_k - f^k (x^s),
\FF^l \left[ \Psi_l \right] - \f{1}{2} \tilde{\Psi}_l \star \Psi_l - 
f^l (x^r) \right]_{\star}
&=& i\theta^{kl},\nonumber\\
\left[ \FF^k \left[ \Psi_k \right] - \f{1}{2} \tilde{\Psi}_k \star 
\Psi_k - f^k (x^s),
p_l \right]_{\star} = i\hbar \delta^{k}_{l}.
\label{constrnoncomm}
\eea
It is a simple exercise to write down the explicit form of these 
relations. They should be used together with the dynamical equation for 
$\FF^k$ which can be obtained by replacing the derivative of the potential 
with respect to $\Psi$ in the Schr\"{o}dinger equation for 
$\tilde{\Psi}_k$ \cite{fm1}. Its explicit form is given by the following relation
\bea
&&\mbox{Symm} \{ \f{1}{2} \left( \FF^{k(1)} - \FF^{k(2)} \right)^{-1} 
\star \left[ \mbox{Symm} \left[ \f{1}{2} \left( \FF^{k(1)} - 
\FF^{k(2)}\right)^{-1}\star \FF^{k(1)}\right]\right] \} (x)
\nonumber\\
&& - \f{2M}{\hbar^2}\{ \left( V_k - E \right) \star \FF^{k(1)}\}(x) = 
0,
\label{dyneqsch}
\eea
where Symm indicates the symmetrized star product with respect to the 
two arguments. The upper index $(i)$ next to the quantum potential 
represents its derivative of the $i$th order with respect to the wave 
function $\Psi_k$.

To conclude, we have derived the relation between the noncommutative 
coordinates in $R^{2N}$ and the wave functions of the algebraic 
Schr\"{o}dinger equation. If the noncommutative wronskian is constant, the 
relevant relations are (\ref{xpsi}), (\ref{constrnoncomm}) and 
(\ref{dyneqsch}). The first of these relations expresses the coordinate - wave 
function duality. The second relation represents a set of constraints on the 
explicit form of the quantum prepotential in terms of solutions of the 
Schr\"{o}dinger equation. These constraints should be enforced on the 
solution of the last equation which is the dual of the Schr\"{o}dinger 
equation in the space of functionals $\FF^{k}$. These results represent 
the generalization of the ones obtained in the commutative case in 
\cite{fm1}. However, in the noncommutative case other possible relations 
are allowed. This can be seen from the equation (\ref{constW}) which 
enforces the linearity in $x^k$ in (\ref{xpsi}). By relaxing this 
condition, the dependence of $x^k$ on $\FF^k$ is modified.

{\bf Acknowledgements}
I am grateful to M. A. Santos for useful discussions. This work was 
supported by the FAPESP Grant 02/05327-3.

\end{document}